\documentclass[reprint,aps,pra,superscriptaddress,longbibliography,floatfix,english]{revtex4-1}

\usepackage{graphicx}
\usepackage{bm, amsmath, amssymb}
\usepackage{color}
 \usepackage{soul}
\usepackage[T1]{fontenc}
\usepackage[latin9]{inputenc}
\usepackage{amstext}
\usepackage{bbold}
\usepackage{esint}
\usepackage{braket}
\usepackage{hyperref}

\newcommand{\sz}[0]{\sigma_z}
\newcommand{\sx}[0]{\sigma_x}
\newcommand{\sy}[0]{\sigma_y}

\def\be{\begin{equation}}
\def\ee{\end{equation}}
\def\bea{\begin{eqnarray}}
\def\eea{\end{eqnarray}}

\usepackage{babel}

\newcommand{\addkm}[1]{{\color{black} #1}}
\newcommand{\revkm}[1]{{\color{black} #1}}
\newcommand{\adddian}[1]{{\color{black} #1}}

\begin{document}

\title{Homodyne monitoring of post-selected decay}

\author{D. Tan}
\affiliation{Department of Physics, Washington University, St.\ Louis, Missouri 63130}
\author{N. Foroozani}
\affiliation{Department of Physics, Washington University, St.\ Louis, Missouri 63130}
\author{M. Naghiloo}
\affiliation{Department of Physics, Washington University, St.\ Louis, Missouri 63130}
\author{A. H. Kiilerich}
\affiliation{Department of Physics and Astronomy, Aarhus University, Ny Munkegade 120, DK-8000 Aarhus C, Denmark}
\author{K. M\o lmer}
\affiliation{Department of Physics and Astronomy, Aarhus University, Ny Munkegade 120, DK-8000 Aarhus C, Denmark}
\author{K. W. Murch}
\affiliation{Department of Physics, Washington University, St.\ Louis, Missouri 63130}
\affiliation{Institute for Materials Science and Engineering, St.\ Louis, Missouri 63130}

\date{\today}

\begin{abstract}
We use homodyne detection to monitor the radiative decay of a superconducting qubit. According to the classical theory of conditional probabilities, the excited state population differs from an exponential decay law if it is conditioned upon a later projective qubit measurement. Quantum trajectory theory accounts for the expectation values of general observables, and we use experimental data to show how a homodyne detection signal is conditioned upon both the initial state and the finally projected state of a decaying qubit. We observe, in particular, how \emph{anomalous} weak values occur in continuous weak measurement for certain  pre- and post-selected states. Subject to homodyne detection, the density matrix evolves in a stochastic manner, but it is restricted to a specific surface in the Bloch sphere. We show that a similar restriction applies to the information associated with the post-selection, and thus bounds the predictions of the theory.
\end{abstract}

\maketitle

\section{Introduction}

\begin{figure}\begin{center}
  \includegraphics[width=0.4\textwidth]{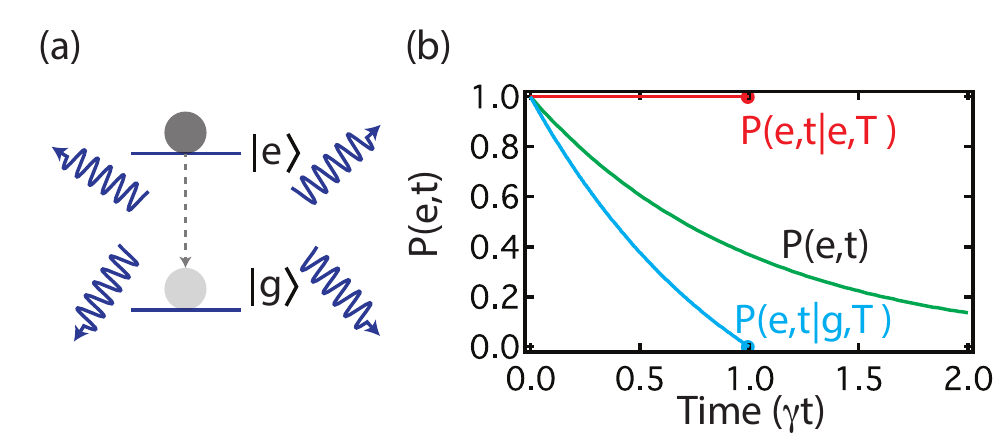}
 \end{center}
  \caption{ \label{fig0}
  Modification of exponential decay with known, classical final states.  (a), We consider a two-level system decaying with a rate $\gamma$. (b), If the system is initialized in the excited state $|e\rangle$ at $t=0$ the probability to find the system in the excited state unconditioned on any later information is given by $P(e,t) = \exp(-\gamma t)$. However, later knowledge of the state at a final time $T$ alters the excited state probability. The red curve shows $P(e,t|e,T)$ and the blue curve shows $P(e,t|g,T)$ for $T=1/\gamma$. }
 \end{figure}

Exponential decay is a fundamental process in classical and quantum physics \cite{ purc46,Asta10}. While the fraction of a large ensemble of systems surviving decay with a rate $\gamma$ until any given time $t$ is represented by an exponential law, $\exp(-\gamma t)$, if the radiative decay of a single system is monitored as a function of time, its actual state evolves in a conditional manner and differs in general from the exponential behavior  \cite{nagh16, camp13, camp16}.  In a similar way to how the state of a quantum system evolves in time subject to information retrieval from measurements, our probabilistic description of a system at a given time in the past is also influenced by information retrieved after that time. To illustrate this, consider how the exponential decay law is modified if we observe the time evolution of a single quantum (or classical) system for which we know the state at a given final time $T$. If an initially excited two-level system, decaying with a rate $\gamma$, is observed to be still in its excited state at time $T$, a previous measurement could \addkm{not possibly} have found the system in its ground state, i.e.,  the exponential decay is replaced by a constant unit excitation probability as illustrated in Fig. \ref{fig0}. If, on the other hand, the system is found in the ground state at the finite time $T$, with what probability would one have found it in the excited state before $T$? This is a simple exercise in conditional probabilities \cite{Stein95}: Let $P(\alpha,t;g,T)$ denote the joint probability that the initially excited system is in state $\ket{\alpha} = \ket{e}$ or $\ket{g}$ at time $t$ \textit{and} in the ground state at time $T$. These joint probabilities can be written in terms of the conditional probabilities, $P(e,t;g,T) = P(e,t)P(g,T|e,t) = e^{-\gamma t} (1-e^{-\gamma(T-t)})$, and $P(g,t;g,T) = P(g,t)P(g,T|g,t) = (1-e^{-\gamma t})\times 1$. The excited state probability at time $t$, conditioned on the initial excited and final ground state at time $T$, is thus given by the ratio,
\begin{equation} \label{eq:classical}
P(e,t|g,T) = \frac{e^{-\gamma t} (1-e^{-\gamma(T-t)}) }{e^{-\gamma t} (1-e^{-\gamma(T-t)}) +(1-e^{-\gamma t})},
\end{equation}
which, as shown in Fig.~\ref{fig0}, interpolates smoothly between unity at $t=0$ and zero at $t=T$.  Equation~\eqref{eq:classical} reflects the predictions we can make about the system state, i.e., the measurement at time $T$ does not impose a physical interaction with the system at time $t$; it merely updates our (present) knowledge about it.

In this article we consider measurements by homodyne detection of the field emitted by a quantum system prepared in an initial state and eventually measured in a given final state. Measurements on quantum systems subject to pre- and post-selection have been subject to theoretical and experimental analysis \cite{ahar10, ahar09, ahar98, ahar91, tan16,tsan09, Guev15, Ryba15} and can be generally described with the Past Quantum State (PQS) formalism \cite{gamm13}.
Here we use this formalism to analyze the outcome of homodyne detection of the signal emitted during spontaneous decay by a qubit system, conditioned on both its initial preparation and on a later projective detection, and we show how anomalous weak values in the continuous measurement signal emerge for certain pre- and post-selected states. We then examine how the initial and final states, together with the continuous measurement record, combine to describe the probability distribution of different qubit observables at any given time:
Supplementing the stochastic evolution of the quantum system conditioned on
the measurement record obtained before a given time $t$ with the information accumulated after time $t$, we observe that the PQS predictions
are at any time confined to certain regions in the Bloch vector picture.
%

This article is organized as follows. In Sec.~\ref{sec:II}, we first describe the experimental setup and present the past quantum state theory.
We then compare our experimental signal arising from homodyne detection of the radiation emitted by the qubit, conditioned on its initial preparation and final projective measurement, to the predictions of PQS theory. In Sec.~\ref{sec:III}, we assess the back-action on the quantum emitter due to homodyne detection of the radiated field, and we determine effective Bloch vector components yielding predictions for projective qubit measurements. Sec.~\ref{sec:IV} concludes the article.

\section{Prediction and retrodiction of the homodyne signal from a decaying two-level emitter} \label{sec:II}
Our experiment is realized in a hybrid two-level system which behaves as a quantum emitter that when initially prepared in the excited state $ |e \rangle$ radiates at its resonant frequency  $\omega_0/2\pi = 6.541$ GHz. The emitter is comprised of a transmon qubit embedded in a 3D aluminum cavity \cite{koch07, Paik11} connected to a 50~$\Omega$  transmission line.
The interaction between the emitter and the transmission line is described by a Hamiltonian $H_{\mathrm{int}}\propto a^\dagger\sigma_-+a\sigma_+$, where $a^\dagger (a)$ is the creation (annihilation) operator for a photon in the transmission line, and  $\sigma_+ (\sigma_-)$ is the pseudo-spin raising (lowering) operator.   The strength of this interaction is given by the Purcell enhanced \cite{purc46_2} radiative decay rate $\gamma =  1.628 $ $\mu$s$^{-1}$ into the transmission line.
We use a near-quantum-limited Josephson parametric amplifier to perform homodyne detection of the \addkm{fluorescence} from the excited $ |e \rangle$ to ground state $ |g \rangle$ transition. \addkm{The homodyne measurement signal is proportional to the amplitude of a specific field quadrature, $a^\dagger e^{i\phi} + a e^{-i\phi}$, and by virtue of the interaction Hamiltonian is a measurement of the corresponding emitter dipole $\sigma_-e^{i\phi} + \sigma_+e^{-i\phi}$. By adjusting the homodyne phase $\phi = 0$, the resulting homodyne signal conveys information about the $\sigma_x$ dipole observable of the qubit \cite{nagh16}}.

Using a classical drive, we may prepare the emitter in an arbitrary initial superposition state. As the qubit decays, the master equation for the density matrix yields the mean dipole and, hence, predicts the mean value of the time dependent emitted signal, as measured by homodyne detection.  If we also know the outcome of a later, projective measurement on the system, the expected outcome of the homodyne measurement changes and is given by the past quantum state theory, which generalizes our introductory classical analysis of joint probabilities to quantum systems.

In quantum mechanics, a general measurement is described by POVMs, i.e., a set of operators $M_m$ satisfying $\sum_{m}M_m^\dagger M_m=\mathbb{1}$. If the system at time $t$ is described by the density matrix $\rho_t$, the probability for outcome $m$ is $P(m)=\mathrm{Tr}(M_m \rho_t M_{m}^\dagger )$, which coincides with Born's rule in the case of projective measurements.

The POVM operators associated with a homodyne fluorescence detection signal $V$, obtained with a detector efficiency $\eta$, are given by \cite{wisebook, jaco06}
\begin{equation} \label{eq:MV}
M_V  = \left(\frac{1}{2\pi\gamma dt}\right)^{\frac{1}{4}} \mathrm{e}^{\frac{-V^2}{4\gamma dt}}\left(1 - \frac{\gamma dt}{2} \sigma_+ \sigma_- + \sqrt{\eta }\gamma\sigma_- V\right),
\end{equation}
and satisfy $\int M_V^\dagger M_V dV = \mathbb{1}$. The probability for the measurement to yield a value $V$ is
\begin{align}\begin{split}
 P(V) &= \mathrm{Tr}( M_V\rho_t M_{V}^\dagger )\\
 &=\frac{1}{\sqrt{2\pi \gamma dt}} \exp{{\left(\frac{-V^2}{2\gamma dt}\right)}} (1+\sqrt{\eta} \gamma \langle \sigma_x\rangle V)
 \\
 &\simeq \frac{1}{\sqrt{2\pi \gamma dt}}\exp\left(-\frac{(V-\sqrt{\eta}\gamma\langle \sigma_x\rangle dt)^2}{2\gamma dt}\right),
 \end{split}
\end{align}
leading to the expected average value, $\overline{V} = \int V P(V) dV=\sqrt{\eta}\gamma\braket{\sigma_x} dt$, with characteristic Gaussian fluctuations.

If the outcome of later measurements on the system are available, they contribute to our knowledge about the system at time $t$ and the resulting modification of the outcome probabilities for the earlier measurement can be written \cite{gamm13}:
\begin{equation}\label{Ppe}
P_p(m,t)=\frac{\mathrm{Tr}(M_m \rho_t M_{m}^\dagger E_t)}{\sum_n\mathrm{Tr} (M_n \rho_t M_{n}^\dagger E_t)},
\end{equation}
where the positive, Hermitian {\it effect matrix} $E_t$ depends on the information accumulated from time $t$ to a final time $T$.

Equation~(\ref{Ppe}) reduces to the classical example offered in the Introduction (Eq.~(\ref{eq:classical})) when the $M_m$ are taken to be the projection operators on the excited and ground states of the emitter, while for homodyne detection, it yields the probability for the measurement signal $V$ conditioned on both prior and posterior measurements,
\begin{equation} \label{Ppv}
P_p(V,t)= \frac{\textrm{Tr}(M_V \rho_t M_V^\dagger E_{t})}{\int dV' \textrm{Tr}(M_{V'} \rho_t M_{V'}^\dagger E_{t})}.
\end{equation}

The density matrix of a decaying quantum system obeys the master equation $d\rho_t= \gamma dt\mathcal{D}[\sigma_-]  \rho_t \equiv \adddian{\gamma dt}\sigma_- \rho_t\sigma_+-\frac{1}{2}\adddian{\gamma dt}\{\sigma_+\sigma_-, \rho_t\}$ with the time dependent solution expressed in terms of the matrix elements of $\rho_t$,
\begin{align} \label{eq:rho-nonoise}
\begin{array}{ll}
\rho^{ee}_t& = \rho^{ee}_0 e^{-\gamma t},
\\
\rho^{ge}_t& = \rho^{ge}_0 e^{-\frac{\gamma}{2}t},
\\
\rho^{gg}_t& = 1-\rho^{ee}_0 e^{-\gamma t}.
\end{array}
\end{align}
Similarly, the matrix $E$ solves the adjoint equation $dE_t= \gamma dt \mathcal{D}^\dagger[\sigma_-] E_t \equiv  \adddian{\gamma dt}\sigma_+ E_t\sigma_- -\frac{1}{2}\adddian{\gamma dt}\{\sigma_+\sigma_-, E_t\}$, where we apply the convention, $dE_t \equiv E_{t-dt}-E_t$, because we shall solve the equation backwards in time.  Equation~(\ref{eq:rho-nonoise}) does not conserve the trace, but this is not a formal problem, since Eqs.~(\ref{Ppe}, \ref{Ppv}) are explicitly renormalized. The (backwards) evolution of $E_t$ from time $T$ yields the solution,
\begin{align}
\begin{array}{ll}\label{eq:E-nonoise}
E^{gg}_t& = E^{gg}_T,
\\
E^{ge}_t& = E^{ge}_T e^{-\frac{\gamma}{2} (T-t) },
\\
E^{ee}_t& = E^{gg}_T+(E^{ee}_T-E^{gg}_T) e^{-\gamma (T-t)},
\end{array}
\end{align}
where $E_T$ is the projection operator on the state of the final heralding measurement (post-selection). In the absence of post-selection, $E_T$ is the identity matrix, and (\ref{eq:E-nonoise}) also yields the identity matrix for all earlier times. \addkm{In this case $P_p(m,t)$, Eqs.\ (\ref{Ppe}, \ref{Ppv}) reduce} to the usual Born rule for quantum expectation values.

\adddian{From Eqs.\ (\ref{Ppv}, \ref{eq:rho-nonoise}, \ref{eq:E-nonoise}), we can express the retrodicted mean value
$\overline{V}_p(t) = \int VP_p(V,t) dV$ of the homodyne signal to first order in the infinitesimal time interval $dt$
by the matrix elements of $\rho_t$ and $E_t$,
\begin{equation}\label{Vp}
\overline{V}_p(t) = \frac{2\sqrt{\eta} \gamma dt\  \mathrm{Re}[E_t^{gg}\rho_t^{eg}+\rho^{ee}_tE^{ge}_t]}{\textrm{Tr}(\rho_t E_t)}.
\end{equation}}

We shall compare Eq.~(\ref{Vp}) with the experimental homodyne detection signal averaged over many experiments, for different choices of the initial state $\rho_0$ and final projection $E_T$. 

\begin{figure}[t]\begin{center}
\includegraphics[angle = 0, width =.5\textwidth]{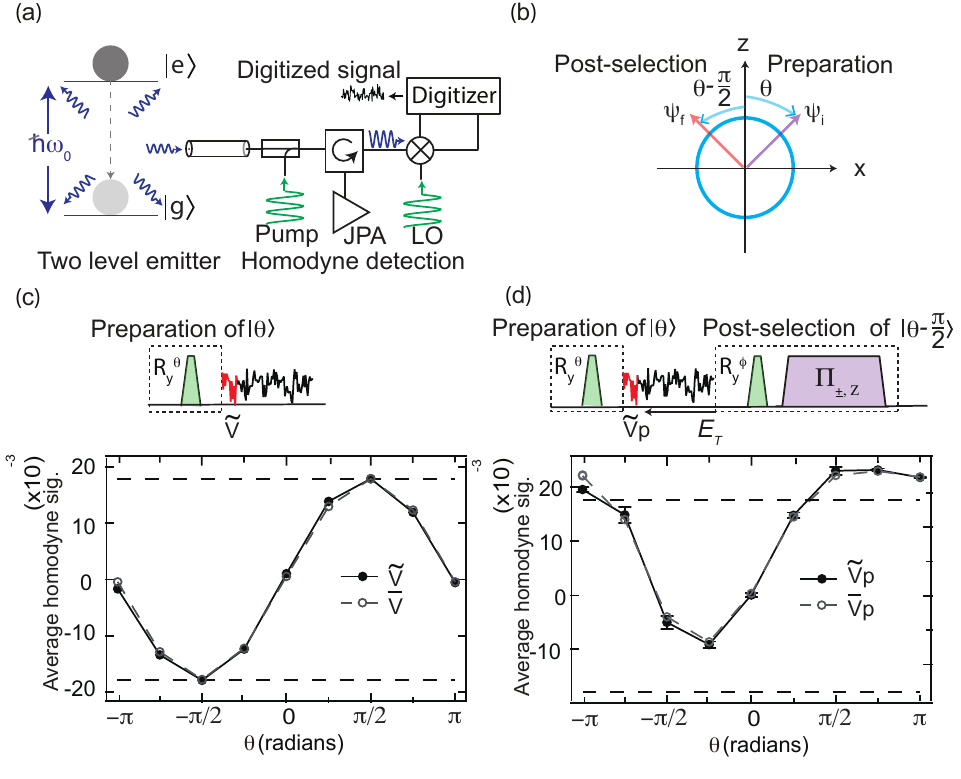}
\end{center}
\caption{ Average homodyne signal. \adddian{(a), The fluorescence from a two-level emitter radiating at frequency $\omega_0$ can be monitored using homodyne detection. (b), Bloch representation of the pre-selection  of the qubit state $ |\psi_i\rangle= |\theta\rangle$ and post-selection in $ |\psi_f\rangle=|\theta-\frac{\pi}{2}\rangle$. (c), The experimental sequence prepares the emitter in a state $ |\theta\rangle$ and we display the average homodyne signal. The solid line with solid dots is the measured average homodyne signal.  The dashed line with hollow dots is the theory predicted mean value $\overline{V}$. More than $5\times10^6$ experimental repetitions are used for each $\widetilde{V}$, leading to a statistical uncertainty of order $4\times 10^{-4}$. (d), The experimental sequence prepares the emitter in the state $ |\theta\rangle$ and post-selects it in the state $ |\theta-\frac{\pi}{2} \rangle$ and we display the average homodyne signal based on $\rho_t$ and $E_t$. The solid line with solid dots is the measured average homodyne signal with pre- and post-selection. The dashed line with hollow dots is the theory predicted mean value $\overline{V}_p$ as calculated in Eq.~\eqref{Vp}. More than $3\times10^4$ experimental repetitions are used for each $\widetilde{V}_p$, leading to a statistical uncertainty of order $5.7\times 10^{-3}$. The error bars indicate the standard deviation associated with the drift in the experimental setup from three repetitions of the experiment.} }\label{Vbar:fig}
\end{figure}

We first examine the experimental average homodyne signal that is obtained without post-selection $\overline{V} = \sqrt{\eta}\gamma\langle \sigma_x\rangle dt$.  We prepare the emitter in a state $\ket{\theta} = \cos\frac{\theta}{2}\ket{g}+\sin\frac{\theta}{2}\ket{e}$ by a rotation pulse $R_y^{\theta}$, and obtain the average homodyne signal $\widetilde{V}$ right after the preparation pulse by integrating $60$ ns of recorded homodyne signal as depicted in Fig. \ref{Vbar:fig}c. In Fig. \ref{Vbar:fig}c, we display our experimental results $\widetilde{V}$, testing the predicted average signal $\overline{V}$ for different $\theta$. $\widetilde{V}$ oscillates as a function of $\theta$ and reaches a maximum (minimum) at $\theta=\frac{\pi}{2}$ ($\theta=-\frac{\pi}{2}$ ) as expected \cite{nagh16}. The experimental and theoretical curves are in good agreement and show that the average homodyne signal $|\widetilde{V}|$ without post-selection never exceeds the maximum value $\sqrt{\eta} \gamma dt$ (dashed horizontal lines).

To confirm the theory prediction for the mean signal with post-selection, $\overline{V}_p$, we conduct the experimental sequence illustrated in Fig.~\ref{Vbar:fig}d. We first initialize the qubit in state $ |\theta\rangle$, then record $0.5\,\mu$s of homodyne signal and finally post-select the state $ |\theta-\frac{\pi}{2} \rangle$. The average, post-selected signal $\widetilde{V}_p$ is obtained by averaging $60$ ns of homodyne signal right after the initial state preparation pulse from the experimental runs which successfully pre-select state $ |\theta\rangle$  and post-select state $ \ket{\theta-\frac{\pi}{2}}$. After correcting for the post-selection fidelity (see Appendix~\ref{sec:AppendixD}) the experimental results $\widetilde{V}_p$ are in good agreement with the theory prediction $\overline{V}_p$, calculated from Eqs.~(\ref{eq:rho-nonoise}, \ref{eq:E-nonoise}, \ref{Vp}) with $\rho_0= \ket{\theta}\bra{\theta}$ and $E_T=|\theta-\frac{\pi}{2} \rangle \langle \theta-\frac{\pi}{2}|$.
Furthermore, we observe anomalous weak values \cite{ahar98} where $|\widetilde{V}_p|$ exceeds $\sqrt{\eta} \gamma dt$.  
This is due to the low overlap between the pre- and post-selected states when $ \theta=\{-\pi, \frac{\pi}{2}, \frac{3 \pi}{4}, \pi\}$ as displayed in Fig.~\ref{Vbar:fig}d.
Note that ideally we could obtain the average $\overline{V}_p$ by post-selecting state $\ket{\theta-\frac{\pi}{2}}$ immediately after the $60\,n$s signal integration,  but transient behavior associated with the rotations and readout affects the homodyne signal. Therefore, as indicated in Fig.~\ref{Vbar:fig}d, we wait for $0.5\,\mu$s before making the post-selection measurement.


\section{Evolution dynamics subject to homodyne detection}\label{sec:III}

In our experiment, \adddian{the emitter state is continuously monitored with the homodyne signal which is  sensitive to the $\sigma_x$ component of the two-level system. If we, rather than averaging over many experiments, consider a single run of the experiment, the state of the system evolves in time as a quantum trajectory which can be inferred from the record of the detected homodyne signal.} Homodyne detection with efficiency $\eta$  gives rise to a signal $V= \sqrt{\eta}\gamma \mathrm{Tr}[\sigma_x\rho_t] dt +  \sqrt{\gamma}dW_t$ with a stochastic Wiener increment $dW_t$ with zero mean and variance $dt$ \cite{Jacbook}, and the density matrix of the emitter solves the stochastic master equation (SME) \cite{wisebook},
\begin{equation}\label{rhoeqn}
 d\rho_t = \gamma dt\mathcal{D}[\sigma_-] \rho_t  + \sqrt{\eta }\left(V-\sqrt{\eta}\gamma\mathrm{Tr}[\sigma_x \rho_t]dt\right) \mathcal{H} [\sigma_- ] \rho_t ,
\end{equation}
where the term proportional to $\mathcal{H}[\sigma_-]\rho_t = \sigma_- \rho_t+\rho_t\sigma_+-\mathrm{Tr}[(\sigma_- +\sigma_+)\rho_t]\rho_t$ is added to the unobserved master equation to account for the stochastic measurement backaction. The trajectory followed by a monitored quantum emitter is well described by the stochastic master equation \eqref{rhoeqn}, and quantum trajectories for the density matrix $\rho_t$ have been studied, e.g., in \cite{camp16, nagh16, murc13traj, webe14, Tan15}.

The homodyne signal $V$ is scaled to have a variance $\sigma^2 = \gamma dt$, and by recording two histograms for $V$ separated by $\Delta V = 2\sqrt{\eta} \gamma dt$, the quantum efficiency of our experimental setup is found to be $\eta=0.3$ (see Appendix~\ref{sec:AppendixB}). Note that this scaling yields a \emph{dimensionless} signal $V$, whereas under other conventions it has units of $(\mbox{time})^{\frac{1}{2}}$ \cite{Bolu14, camp16}.

Similar to the density matrix $\rho_t$ being now conditioned on the initial state and the homodyne detection record until time $t$, the matrix $E_t$ at time $t$ is conditioned on the homodyne signal recorded after $t$. It solves the adjoint counterpart of the (SME) Eq.\ (\ref{rhoeqn}) backwards from the final time $T$ \cite{gamm13},
\begin{equation}\label{Eeqn}
dE_t = \gamma dt \mathcal{D}^\dagger[\sigma_-] E_t  + \sqrt{\eta } \left(V-\sqrt{\eta}\gamma\mathrm{Tr}[\sigma_x E_t]dt\right)\mathcal{H} [\sigma_+ ] E_t.
\end{equation}

\subsection{Bloch representation of $\rho_t$ and $E_t$}
\label{sec:BlochRepresentation}

To graphically present the results, the density matrix of a two-level quantum system may be represented by a real Bloch vector $(x_\rho,y_\rho,z_\rho)$,
\begin{align}
\rho_t = \frac{1}{2}\left(\mathbb{1}+x_\rho\sx+y_\rho\sy+z_\rho\sz\right),
\end{align}
where $u_\rho = \mathrm{Tr}(\sigma_u\rho_t)$ for $u =x,y,z$.
The stochastic master equation \eqref{rhoeqn} describes how the evolution of a decaying qubit monitored by homodyne detection is conditioned on the measurement signal. For perfect detection ($\eta=1$), an initially pure state remains pure and the Bloch vector explores the surface of the unit Bloch sphere, while imperfect detection ($0\leq\eta<1$) leads to a mixed state \textit{inside} the Bloch sphere.
In our experiments, the system is prepared (and post-selected) in states with vanishing $\langle \sigma_y \rangle$, and as the homodyne detection effectively probes the $\sigma_x$ operator; the (conditional) $y_\rho$ component of the Bloch vector remains zero at all times.
The SME is, hence, equivalent to the following coupled stochastic equations for the $x_\rho$ and $z_\rho$ Bloch vector components,
\begin{align}\label{eq:rhoBloch}
\begin{split}
dx_\rho &=-\frac{\gamma}{2} x_\rho dt + \sqrt{\eta} (1\adddian{-}z_\rho-x_\rho^2)(V-\sqrt{\eta}\gamma x_\rho dt),
\\
dz_\rho &= \gamma(1\adddian{-}z_\rho)dt-\sqrt{\eta}(1\adddian{-}z_\rho)x_\rho (V-\sqrt{\eta}\gamma x_\rho dt).
\end{split}
\end{align}

Similarly, we wish to introduce a Bloch sphere representation to illustrate the conditional evolution of $E_t$. While the role of $E_t$ in predicting measurement outcomes does not require unit trace due to the normalization factor in Eq.~\eqref{Ppe}, the Bloch sphere representation assumes a normalized state matrix.
\adddian{The term $\mathcal{D}^\dagger[\sigma_-] E_t$ in the SME Eq. \eqref{Eeqn}  is not trace preserving}, but since $\gamma dt\mathrm{Tr}(\mathcal{D}^+[\sigma_-] E_t) = \gamma dt\mathrm{Tr}(\sz E_t)$, we may introduce the following normalized version of the SME,
\begin{align}\label{EeqnNormalized}
\begin{split}
dE_t = \gamma dt \mathcal{D}^\dagger[\sigma_-] E_t-\gamma dt\mathrm{Tr}[\sz E_t]E_t+
\\
\sqrt{\eta } \left(V-\sqrt{\eta}\gamma\mathrm{Tr}[\sigma_x E_t]dt\right)\mathcal{H} [\sigma_+ ] E_t,
\end{split}
\end{align}
and an associated Bloch vector
\begin{align}\label{eq:EBlochRepr}
E_t = \frac{1}{2}\left(\mathbb{1}+x_E\sx+y_E\sy+z_E\sz\right),
\end{align}
\adddian{where $u_E = \mathrm{Tr}(\sigma_u E_t)$ for $u =x,y,z$.}
Just like \eqref{eq:rhoBloch}, we  can obtain a set of stochastic Bloch equations for $E_t$,
\begin{align}\label{eq:EBloch}
\begin{split}
dx_E &=-\frac{\gamma}{2} x_E\left[1\adddian{+}2 z_E+\eta (1\adddian{+}z_E-x_E^2)\right] dt
\\
&+\sqrt{\eta} (1\adddian{+}z_E-x_E^2)V,
\\
dz_E &= -\gamma (z_E\adddian{+}z_E^2\adddian{-}\eta(1\adddian{+}z_E)x_E^2) dt
\\
&-\sqrt{\eta}(z_E\adddian{+}1)x_E V.
\end{split}
\end{align}

\begin{figure}\begin{center}
\includegraphics[angle = 0, width =0.48\textwidth]{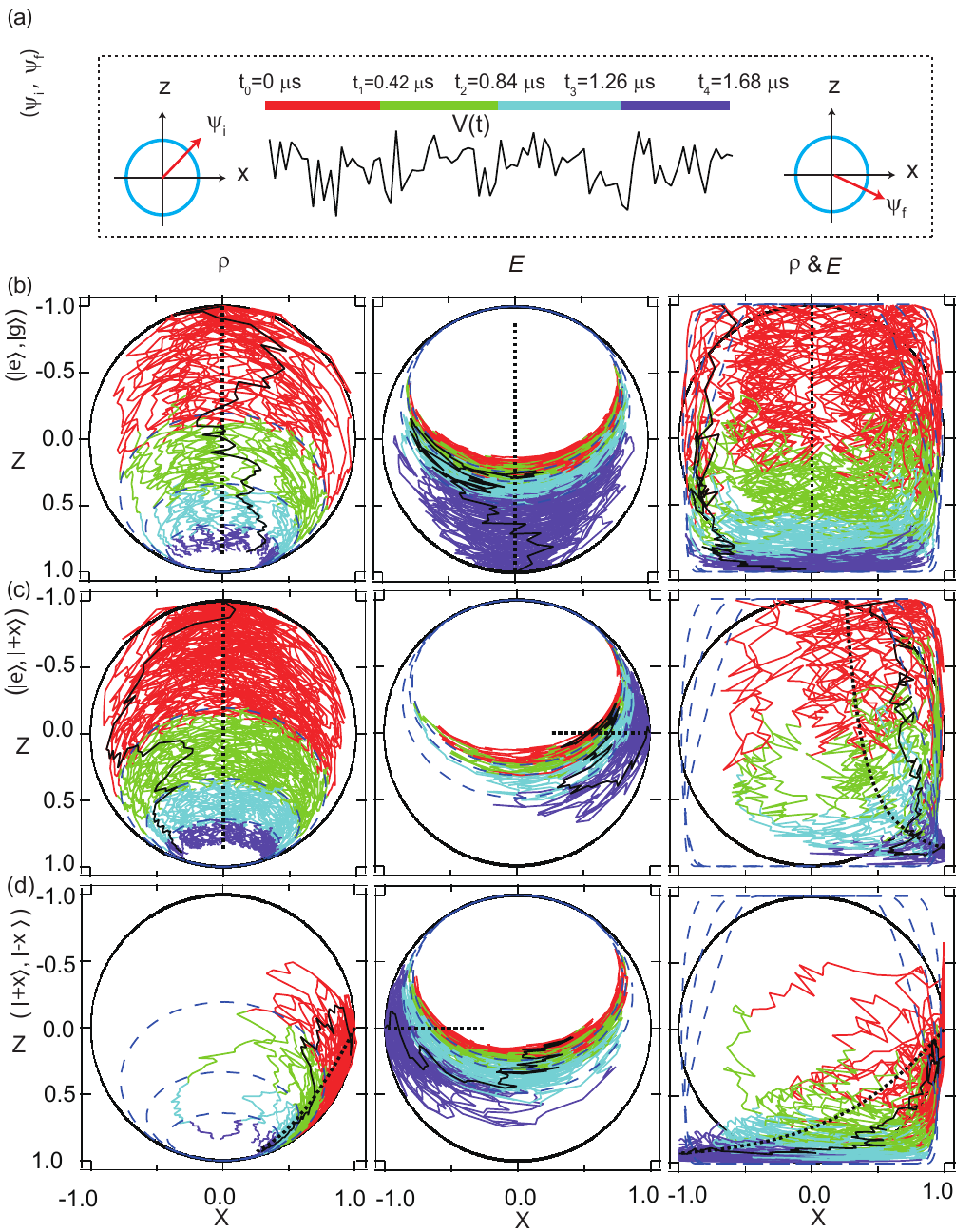}
\end{center}
\caption{Quantum trajectories for pre- and post-selected states. (a), We prepare the qubit in state $|\psi_i\rangle$, the qubit evolves for 1.68 $\mu$s and is post-selected in the state $|\psi_f\rangle$. (b, c, d), Bloch vector representation of the solutions of the stochastic master equations. The three columns show the evolution of the Bloch vector $x$ and $z$ components of $\rho_t$, $E_t$ and of the ($\rho_t$,~$E_t$) retrodiction. The black dashed lines in the figures show the deterministic evolution in the absence of homodyne detection, while the black solid lines represent a single stochastically diffusing trajectory of the emitter. The blue, dashed ellipses in the first and second columns illustrate the deterministic curves to which the stochastically evolving Bloch sphere trajectories are confined at any given time. These combine to yield a restriction of the area explored by the retrodicted expectation values.
The closed blue, dashed curves in the third column mark the outer boundary of this area.
}
\label{fig5}
\end{figure}

In Fig.\ \ref{fig5}a, we show schematically how we prepare the emitter in the state $ |\psi_i\rangle$, then digitize the detected homodyne signal $V(t)$, accumulated for a time interval of 1.68 $\mu$s and finally measure the emitter in the $ |\psi_f\rangle$ state by a high fidelity projective measurement. Using Eqs.~(\ref{eq:rhoBloch}, \ref{eq:EBloch}), we determine the conditional Bloch vectors for $\rho_t$ and $E_t$ and in Fig.\ \ref{fig5} we show the resulting trajectories with the colors red, green, cyan, and blue, corresponding to different time intervals $[0.42 n,\ 0.42 (n+1)]\  \mu$s, $(n=0,1,2,3)$. Here the panels b-d represent different choices for the initial and final states. The trajectories for $\rho_t$, shown in the first column in Fig. \ref{fig5}b-d diffuse through the Bloch sphere, but are confined to different deterministic curves for different evolution times (blue dashed lines in Fig. \ref{fig5}) \cite{camp16, nagh16}. In a similar way, the trajectories for $E_t$ diffuse backwards in time through the Bloch sphere from the post-selected state and they are also, for different evolution times, confined to different deterministic curves. Analytic expressions for these curves are provided in the following subsection.

The Bloch vector components of $\rho_t$ are the expectation values of the Pauli operators, but can also be written as the weighted mean value of their eigenvalues, e.g.,
$\langle \sigma_z \rangle = \rho_t^{gg}\cdot 1 + \rho_t^{ee} \cdot (-1)= 2\rho_t^{gg}-1$. Since the Past Quantum State answers the question: "What is the probability that a measurement of an observable gave a certain outcome a time $t$?", we can use Eq.~(\ref{Ppe}) to obtain such probabilities for projection operators on the eigenstates of $\sigma_x, \sigma_y$ and $\sigma_z$, and subsequently display the weighted eigenvalues as Bloch vector components, e.g.,
\begin{align}\label{eq:szP}
\braket{\sz}_{p}(t) = P_{p}(\sz=+1,t)-P_{p}(\sz=-1,t).
\end{align}
Similar equations apply for $\sx$ and $\sy$ measurements, and using the Bloch vector representation of $\rho_t$ and $E_t$, the retrodicted expectation values of the three spin components acquire the elegant form,
\begin{align}\label{eq:sxsyszPQS}
\begin{split}
\braket{\sx}_{p}&= \frac{x_\rho+x_E}{1+x_\rho x_E},
\\
\braket{\sy}_{p}&= \frac{y_\rho+y_E}{1+y_\rho y_E},
\\
\braket{\sz}_{p}&= \frac{z_\rho+z_E}{1+z_\rho z_E}.
\end{split}
\end{align}
These expressions are used together with the solutions of Eqs.~(\ref{eq:rhoBloch}, \ref{eq:EBloch}) to plot the trajectories of the retrodicted expectation values in the third columns in Fig.~\ref{fig5}b-d. These trajectories diffuse through the state space, and notably assume values that are outside of the Bloch sphere. This is as expected, since, e.g., the prediction for the outcome of a measurement of the ground state population at late times is unity with almost certainty, while post-selection upon a final measurement of $\sigma_x$, certifies that an immediately foregoing measurement of $\sigma_x$ would have to yield the same result. Note that this is not at variance with Heisenberg's uncertainty relation which concerns only predictions of future measurements and does not apply for the combined prediction and retrodiction of observations. We emphasize that, while the mean values and probabilities for the outcome of measurements along any rotated spin direction simply follow from the projection of the Bloch vector along those directions, due to the non-linear expressions in Eq.\ (\ref{eq:sxsyszPQS}) the same reasoning does not apply for the vector plotted in the third columns in Fig.~\ref{fig5}b-d. Prediction of the spin measurement along a 45 degree direction between the $x$ and $z$ axes, would require a separate calculation, using the $\rho_t$ and $E_t$ Bloch vector components along that direction.

\subsection{Deterministic properties of  $\rho_t$ and $E_t$}
\label{sec:detell}

In this section, we examine the character of the stochastic trajectories in more detail.
In Ref.\ \cite{camp16} it is derived how the stochastic evolution of a decaying qubit subject to heterodyne detection is at all times confined to the surface of a deterministic spheroid inside the Bloch sphere. In the case of homodyne detection,
only one component of the pseudo-spin is probed, and the three dimensional spheroid is replaced by a two dimensional ellipse. Here we wish to extend these results to the Bloch representation Eq.~\eqref{eq:EBlochRepr} of the matrix $E_t$.

\begin{figure}
\centering
\includegraphics[trim=0 0 0 0,width=0.45\textwidth]{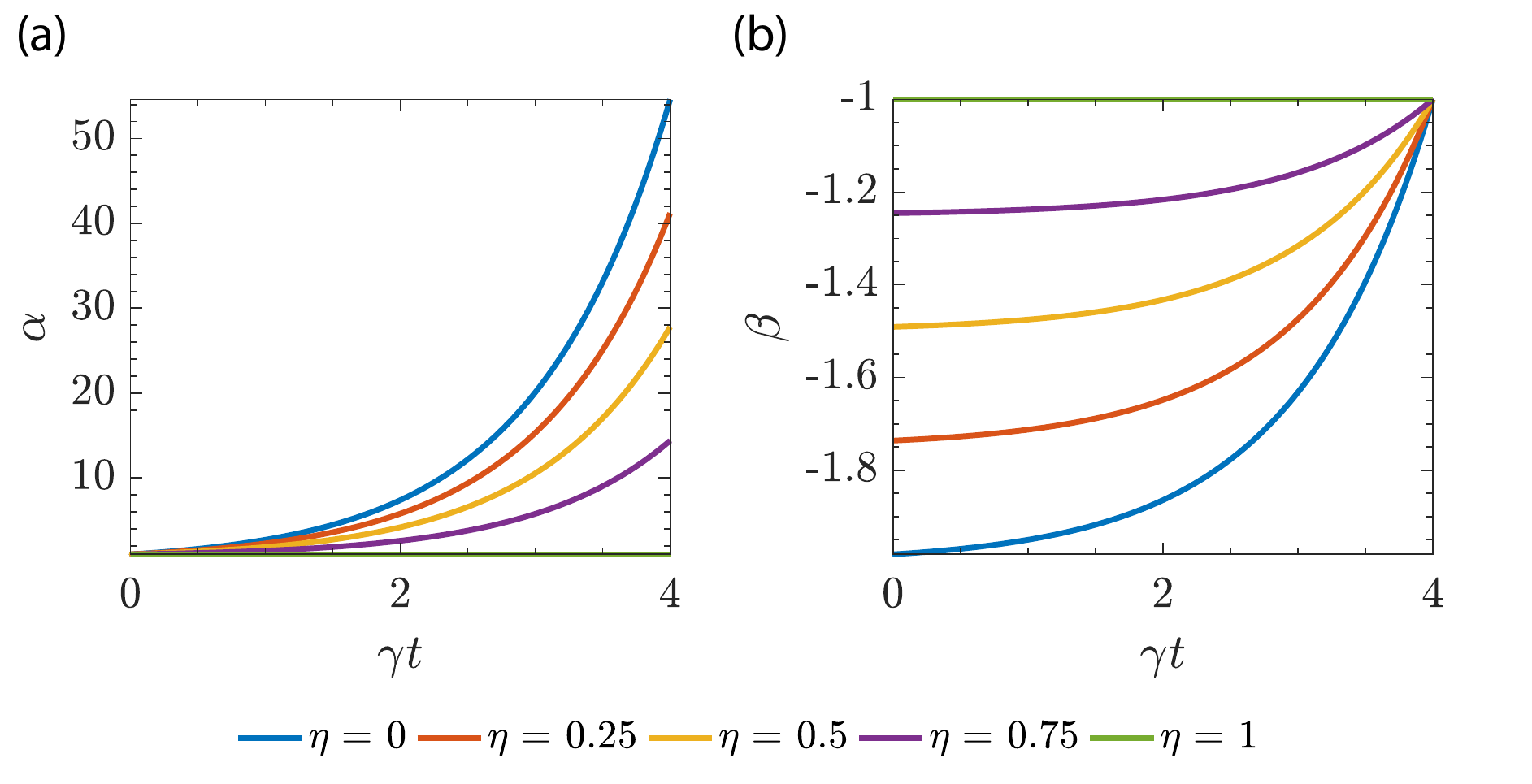}
\caption{Time evolution of functions parametrizing determinstic ellipses. (a), Time evolution given by Eq.~\eqref{eq:at} of the function $\alpha(x_\rho,z_\rho)$ parametrizing the deterministic ellipse \eqref{eq:alphaEllipse} in the Bloch sphere on which a decaying spin subject to homodyne detection is confined.
 (b), Time evolution Eq.~\eqref{eq:bt} of the function $\beta(x_E,z_E)$ for a similar ellipse \eqref{eq:betaellipse} pertaining to the effect matrix and assuming post-selection at time $T=4\gamma^{-1}$ in a pure state.
}
\label{fig:ab}
\end{figure}

For completeness, we \addkm{first  re-derive} the expressions for the ellipse pertaining to the density matrix. The quest is to identify a function $\alpha(x_\rho,z_\rho)$ of the stochastically evolving Bloch components, for which the equation of motion is deterministic. We shall see that such a function exists
and that it indeed describes an ellipse in $(x_\rho,z_\rho)$.
For a generic function, the equation of motion is derived from the stochastic Bloch equations \eqref{eq:rhoBloch},
\begin{align}\label{eq:dalpha}
\begin{split}
d\alpha &= \frac{\partial\alpha}{\partial x_\rho}dx_\rho+\frac{\partial\alpha}{\partial z_\rho}dz_\rho
\\
&+\frac{1}{2}\left[\frac{\partial^2\alpha}{\partial x_\rho^2}(dx_\rho)^2+\frac{\partial^2\alpha}{\partial z_\rho^2}(dz_\rho)^2
+2\frac{\partial^2\alpha}{\partial x_\rho\partial z_\rho}dx_\rho dz_\rho\right],
\end{split}
\end{align}
where the second order terms yield contributions from the noise terms in Eq.~\eqref{eq:rhoBloch} of the same order in $dt$ as the first order deterministic terms.
The evolution of $\alpha(x_\rho,z_\rho)$ is deterministic if all terms proportional to $dW_t$ in $d\alpha$ cancel. After applying Eq.~\eqref{eq:rhoBloch} in
Eq.~\eqref{eq:dalpha} this requirement dictates the following form of $\alpha(x_\rho,z_\rho)$,
\begin{align}\label{eq:alp}
\alpha(x_\rho,z_\rho) = \frac{2}{1\adddian{-}z_\rho}-\frac{x_\rho^2}{(1\adddian{-}z_\rho)^2}.
\end{align}
This can be rewritten
\begin{align}\label{eq:alphaEllipse}
\alpha^2(1\adddian{-}z_\rho-1/\alpha)^2+\alpha x_\rho^2=1,
\end{align}
which shows that the Bloch components of $\rho_t$ are at all points in time restricted to an ellipse centred at $(x,z) = (0,1\adddian{-}1/\alpha)$ and with major axis  $1/\sqrt{\alpha}$ ($x_\rho$-direction) and minor axis  $1/\alpha$ ($z_\rho$-direction).
Furthermore, applying Eq.~(\ref{eq:alp}) on the right hand side of Eq.~\eqref{eq:dalpha}
yields an ordinary differential equation for the time evolution of the parametrizing function $\alpha(x_\rho,z_\rho)$,
\begin{align}
\frac{d\alpha}{dt} = \gamma\left(\alpha-\eta\right),
\end{align}
with the solution
\begin{align}\label{eq:at}
\alpha(t) = \eta+[\alpha(t=0)-\eta]\mathrm{e}^{\gamma t},
\end{align}
where $\alpha(t=0)$ follows from Eq.~\eqref{eq:alp} with the initial Bloch components at time $t=0$ .
For any pure initial state $\alpha(t=0)=1$, and the ellipse Eq.~\eqref{eq:alphaEllipse} is the full Bloch sphere.
The time evolution of $\alpha$ for an initial pure state is shown \adddian{in} Fig.~\ref{fig:ab}a for different values of the detection efficiency $\eta$.
$\alpha(t)$ increases, and hence the center $z$-coordinate of ellipse \addkm{increases} with a rate $\gamma$ in accordance with the decay of the qubit.
As a signature of the loss of information associated with non-perfect monitoring,
the axes of the ellipse reduce faster for smaller values of the detector efficiency $\eta$ and  the qubit explores a range of mixed states. At large times, both axes of the ellipse diminish and the (pseudo-)spin is certain to be found in the ground state.

To derive a similar result for $E_t$, we define a generic function
$\beta(x_E,z_E)$ of the Bloch components in Eq.~\eqref{eq:EBlochRepr}, and we seek a form of this function evolving in a deterministic manner.
The equation of motion for $\beta(x_E,z_E)$ follows from the stochastic Bloch equations \eqref{eq:EBloch} in a way equivalent to that for $\alpha(x_\rho,z_\rho)$ in Eq.~\eqref{eq:dalpha}, and the requirement that all terms proportional to $V$ cancel yields
\begin{align}\label{eq:beta}
\beta(x_E,z_E) = \adddian{-}\frac{2}{z_E\adddian{+}1}+\frac{x_E^2}{(z_E\adddian{+}1)^2}.
\end{align}
Rewriting and noting that $\beta<-1$ reveals that $\beta$ parametrizes an ellipse in $(x_E,z_E)$,
\begin{align}\label{eq:betaellipse}
1=\beta^2(z_E\adddian{+}1\adddian{+}1/\beta)^2-\beta(x_E^2+y_E^2),
\end{align}
centered at $(x,z) = (0,\adddian{-}(1+1/\beta))$ and with major axis $1/\sqrt{-\beta}$ ($x_E$-direction) and minor axis $1/\beta$ ($z_E$-direction).

In addition, one finds that the time evolution of $\beta(x_E,z_E)$ fulfills the differential equation,
\begin{align}
\frac{d\beta}{dt}  = \gamma(-\beta+\eta-2),
\end{align}
which must be solved backwards in time from the final value at the time of post-selection $\beta(t=T)$. This gives the following evolution for $\beta(x_E,z_E)$,
\begin{align}\label{eq:bt}
\beta(t) = \eta-2+[\beta(T)-\eta+2]\mathrm{e}^{\gamma (t-T)}.
\end{align}
Equation~(\ref{eq:beta}) provides the value of $\beta(t=T)$ from the Bloch components of the post-selected state. For any pure state $\beta(t=T)=-1$ and the ellipse is the full Bloch sphere. Without post-selection $E_{T} = \mathbb{1}/2$ so $\beta(t=T)=-2$ and the final ellipse is smaller and includes the origin $(x,z) = (0,0)$.
The time evolution of $\beta$ for a final post-selection in a pure state  at time $T=4\gamma^{-1}$ is shown in Fig.~\ref{fig:ab}b for different values of the detection efficiency $\eta$. Similarly to the case of the density matrix, lower efficiency causes a faster (backwards) decay of the ellipse towards that corresponding to a fully mixed effect matrix.

The retrodicted expectation values of the spin components at any point in time during an experiment follows in Eq.~\eqref{eq:sxsyszPQS} from the Bloch components of the density and effect matrices at that point in time. Combining the restriction of $\rho_t$ and $E_t$ to deterministic curves in the Bloch sphere plot, the retrodiction for  $\sigma_x$ and $\sigma_z$ measurements becomes confined to time dependent $(x,z)$ domains. For different realizations of the homodyne signals, the retrodicted outcomes of measurements of the two spin components explore this area, as illustrated in the third columns in Fig.~\ref{fig5}b-d.

\section{Conclusion}\label{sec:IV}
\revkm{We have applied the Past Quantum State formalism} to a quantum emitter continuously monitored by homodyne detection. Our analysis shows how post-selection leads to a modification of the prediction for the emitter state in a simple way that can be understood with a classical analysis of joint probabilities.  The Past Quantum State formalism makes predictions for the average homodyne signal given specific pre- and post-selected states. We have experimentally confirmed these predictions and furthermore observed anomalous weak values in the homodyne signal as is expected for pre- and post-selected states with small overlap. These weak values have been shown to offer metrological advantages under some circumstances \cite{Dixon09, Pang14, Wise02}. By employing quantum trajectory theory, we have studied the conditioned evolution of the emitter state in the Bloch vector representation and presented  trajectories for the predicted and retrodicted state evolution.  These trajectories evolve stochastically, but they are confined to deterministic regions in the Bloch sphere.

\begin{acknowledgements}
A. H. K. and K. M. acknowledge financial support from the Villum Foundation.
A. H. K. further acknowledges support from the Danish Ministry of Higher  Education  and  Science and K.W.M. acknowledges support from the Sloan Foundation. Experimental work was supported by NSF (grant PHY-1607156) and the John Templeton Foundation. This research used facilities at the Institute of Materials Science and Engineering at Washington University.
\end{acknowledgements}

\section*{appendix}\label{sec:appendix}
\appendix

\section{Sample fabrication and parameters}
\label{sec:AppendixA}
\addkm{The experimental set-up is similar to that of our previous work \cite{caus16}. The transmon circuit was fabricated from double-angle-evaporated aluminum on a silicon substrate and  is characterized by charging energy $E_C/h$ = 300 MHz and Josephson energy $E_J/h$ =19.73 GHz.  The circuit was placed at the center of a 3D aluminum waveguide cavity with $\omega_c$ = 7.257 GHz (dimensions $34.15 \times 27.9 \times 5.25$ mm$^3$) which was machined from 6061 aluminum.  The near resonant interaction between the qubit and the cavity is characterized by coupling rate $g/2\pi = 130$ MHz and produces hybrid states.  We use the lowest energy transition $\omega_\mathrm{q}/2\pi = 6.541$ GHz as the quantum emitter.   The emitter coherence properties, $T_1$ = $614$  ns, $T_2^*$ = 800 ns  are measured using standard techniques.  The  Josephson parametric amplifier consists of a 1.5 pF capacitor shunted by a SQUID loop which is composed of two Josephson junctions with critical current $I_0=1\ \mu$A. The amplifier is operated with small threading the SQUID loop and produces 20 dB of gain with an instantaneous 3-dB-bandwidth of 50 MHz.}

\section{Calibration of homodyne signal}
\label{sec:AppendixB}
\begin{figure}\begin{center}
\includegraphics[angle = 0, width =.5\textwidth]{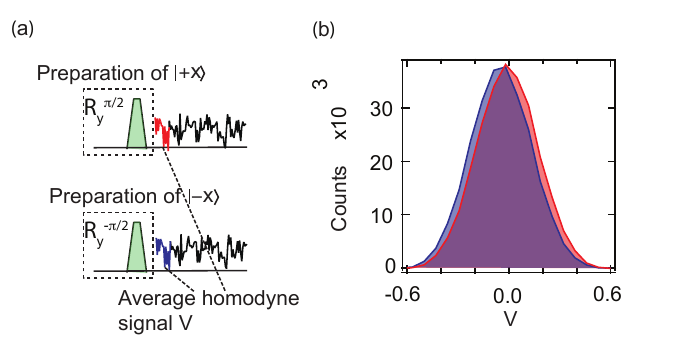}
\end{center}
\caption{Calibration of the homodyne signal. (a), Experimental sequences to prepare the emitter in the $|+x\rangle$ and $|-x\rangle$ state. (b),  Histograms of the homodyne signal for the $|+x\rangle$ state (red) and $|-x\rangle$ state (blue).}\label{PQS:signal}
\end{figure}

We conduct a simple experiment illustrated in Fig.~\ref{PQS:signal}a to calibrate our measurement homodyne signal by applying a $R_{y}^{\frac{\pi}{2}} $($R_{y}^{-\frac{\pi}{2}}$) pulse to prepare the emitter in $|+x\rangle$ or ($|-x\rangle$).  We collect $20$ ns of homodyne signal immediately after the state preparation. We scale the measurement homodyne signal so that the variance $\sigma^2 = \gamma dt$, where the time step  $dt=20$ ns.
From $5\times 10^5$ experimental runs, we obtain
two histograms shown in Fig.~\ref{PQS:signal}b with Gaussian distributions centered at $\pm\sqrt{\eta} \gamma dt$ and separated by $\Delta V = 2\sqrt{\eta} \gamma dt$. Hence, the quantum efficiency of this experiment setup $\eta=0.3$.

\section{High fidelity post-selection measurements}
\label{sec:AppendixC}
\begin{figure}\begin{center}
\includegraphics[angle = 0, width =1\columnwidth]{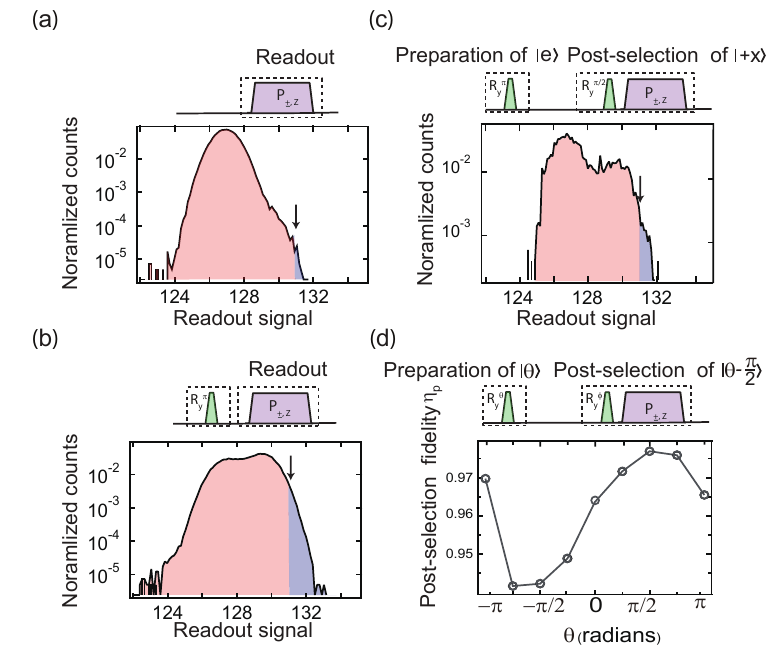}
\end{center}
\caption{Post-selection fidelity. (a), Histogram of the readout when the emitter is in the ground state. The arrow indicates the threshold value for the readout and the blue region indicates the number of error detections. (b), When the emitter is prepared in the excited state, more post-selections are successful. The histograms in (a) and (b) can be used to determine the post-selection error in our experiment. (c), Histogram of readout  when the emitter is prepared in the excited state and post-selected in the $|+x\rangle$ state. (d), The post-selection fidelity for each pair of pre- and post-selected states ($ |\theta\rangle$, $ |\theta-\frac{\pi}{2} \rangle$)}\label{Occ:fig6}
\end{figure}

Post-selection experiments often look for rare events, and in experiments with modest measurement fidelity, post-selection errors can easily contaminate the measurement results.  Here, we focus on maximizing the post-selection fidelity at the expense of the the post-selection efficiency. In our experiment, we realize high fidelity post-selections by adjusting the readout power to the extent that minimizes the error occurrence while maintaining a modest success rate. In the language of photo-detection we want to minimize the dark counts (the post-selection errors) even at the expense of low photo-detection efficiency. We define the post-selection fidelity as the fraction of correct post-selections. We test the post-selection error rate by preparing the qubit in the ground state and then performing a readout measurement. On average, by choosing an appropriate threshold (arrows in Fig.~\ref{Occ:fig6}), we found 2 error occurrences out of 5000 runs of the experiment as shown in the blue region in Fig.~\ref{Occ:fig6}a. If we prepare the qubit in the excited state, we have 314 occurrences from the same number of runs with the same threshold.  At this point, we know the post-selection error rate is below 1\%. To apply this post-selection technique to other states, we simply apply a qubit rotation before the readout.

While it is possible to reduce the post-selection error rate below 1\%, when post-selecting on rare events with for example an expected occurrence of one in $10^{6}$, these post-selection errors will dominate the experimental results. This limits the types of post-selections that can be reliably made, and we focus on post-selections where successes rate (the ratio of number of successful runs to total experiment runs), greatly outweighs the error rate. Fig. \ref{Occ:fig6}d characterizes the post-selection fidelity for different pre- and post-selected states. To test the post-selection fidelity, we conduct the experimental sequence as illustrated in Fig. \ref{Occ:fig6}d; we first apply a \addkm{rotation to} prepare the qubit in the  state $ |\theta \rangle$ in the $x$-$z$ plane of the Bloch sphere. After 0.5 $\mu$s  we then post-select the $ |\theta-\frac{\pi}{2} \rangle$ state by applying a corresponding rotation $R_{y}^{\phi}$ and a projective measurement $\Pi_{\pm, z}$.  \addkm{The post-selection fidelity is the ratio of correct post-selections to the total number of detection events.  The number of incorrect post-selections is the product of the error rate and the number of trials.}

\section{Correction of the predicted mean value $\overline{V}_p$ due to post-selection fidelity}
\label{sec:AppendixD}
In the experiment, we prepare the emitter in the state $ |\theta\rangle=\cos(\frac{\theta}{2})|g\rangle + \sin(\frac{\theta}{2})|e\rangle$ at $ t=0$ and post-select state $ |\theta-\frac{\pi}{2} \rangle=\cos(\frac{\theta-\frac{\pi}{2}}{2})|g\rangle + \sin(\frac{\theta-\frac{\pi}{2}}{2})|e\rangle$ at $t=T$. Ideally, we have the density matrix $\rho_{t=0}(\theta)=|\theta\rangle\langle\theta|$ and the effect matrix $E_{t=T}(\theta)=|\theta-\frac{\pi}{2} \rangle\langle \theta-\frac{\pi}{2} |$.
In the experiment, however, the post-selection fidelity $\eta_p$ is \addkm{sub-unity} as shown in Fig.~\ref{Occ:fig6}d. To account for this in the analysis, the effect matrix $E_T$ at time $T$ for calculating $\overline{V_p}$  is corrected in the following way,
\[E^{gg}_T(\theta)=(1-\eta_p) \cos^2\left(\frac{\theta-\frac{\pi}{2}-\pi}{2}\right)+\eta_p \cos^2\left(\frac{\theta-\frac{\pi}{2}}{2}\right)\]
\[E^{ge}_T(\theta)=\frac{1}{2}(1-\eta_p) \sin\left(\frac{\theta-\frac{\pi}{2}-\pi}{2}\right)+\frac{1}{2} \eta_p \sin\left(\frac{\theta-\frac{\pi}{2}}{2}\right)\]
\[E^{ee}_T(\theta)=(1-\eta_p) \sin^2\left(\frac{\theta-\frac{\pi}{2}-\pi}{2}\right)+\eta_p \sin^2\left(\frac{\theta-\frac{\pi}{2}}{2}\right)\]
After taking the post-selection fidelity $\eta_p$ into consideration, the experimental and theoretical curves agree well as displayed in Fig. \ref{Vbar:fig}d of the main text.

\begin{figure*}[]
\centering
\includegraphics[width=0.7 \textwidth]{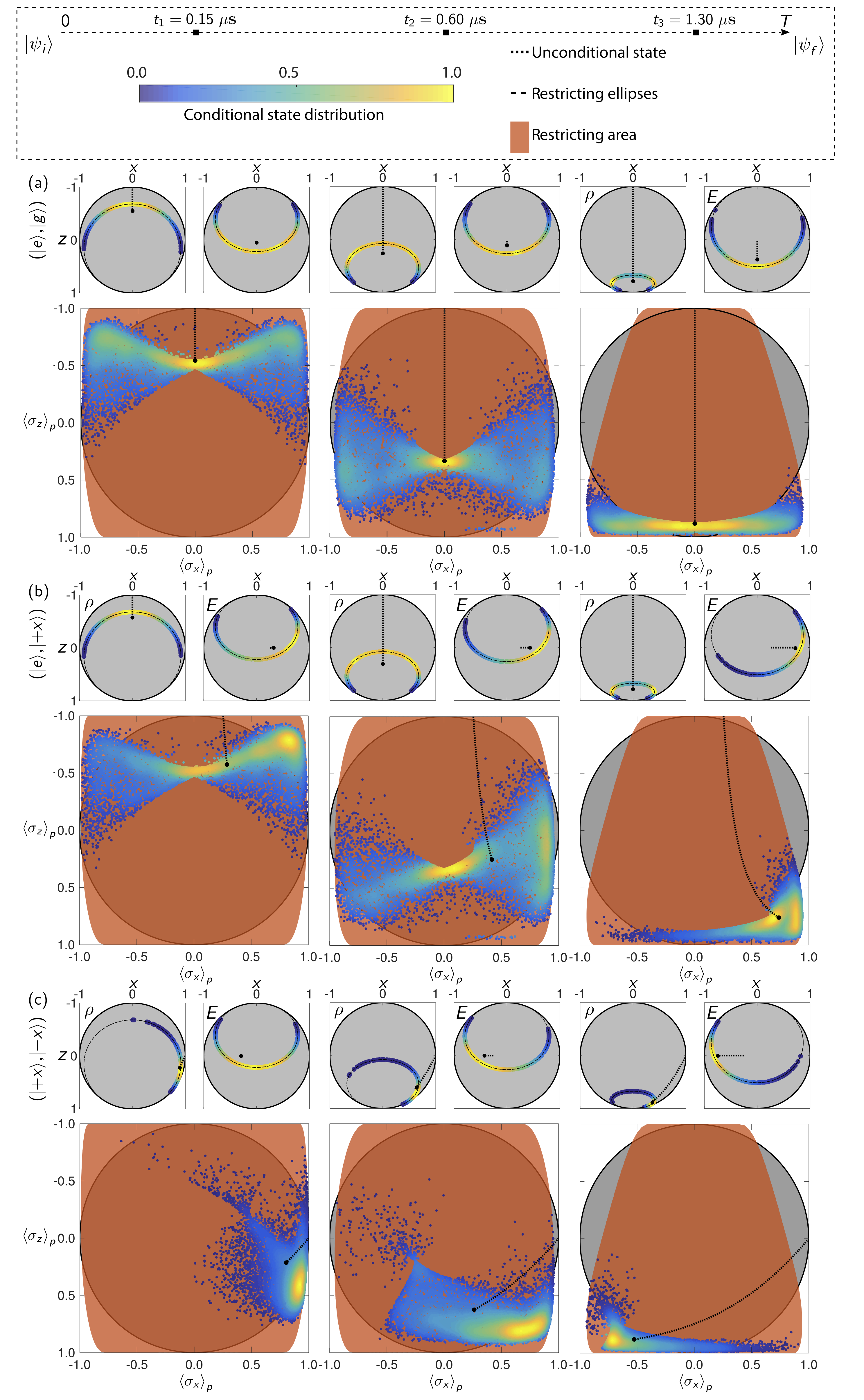}
\caption{Monte Carlo sampled distribution of trajectories.
The (pseudo)-spin is prepared in state $\ket{\psi_i}$, evolves for $T=1.68$ $\mu$s and is post-selected in the state $\ket{\psi_f}$.
(a, b, c), Shown as coloured dots is the distribution the Bloch components of $\rho_t$, $E_t$ and of the retrodicted expectation values $(\braket{\sx}_p,\braket{\sz}_p)$ sampled from $10\,000$ Monte Carlo simulated trajectories with the experimental parameters $\gamma = 1.628$ $\mu$s$^{-1}$ and $\eta=0.3$.
The colors of the dots represent the local density normalized to unity maximum.
 The deterministic ellipses, Eqns.~(\ref{eq:alphaEllipse}) and (\ref{eq:betaellipse}) restricting $\rho_t$ and $E_t$ respectively, are shown as dashed lines in the upper panels. The corresponding deterministic area restricting $(\braket{\sx}_p,\braket{\sz}_p)$ is coloured orange in the lower panels.
For reference, the density and effect matrices and the retrodicted expectation values for an unmonitored system are tracked by dotted lines.
The three columns correspond to different times $t_1 = 0.15$, $t_2= 0.60$  and $t_3=1.30$ $\mu$s.
}
\label{fig:simulations}
\end{figure*}

\section{Distribution of Bloch vector components}
\label{sec:AppendixE}
In this appendix, we revisit the deterministic ellipses and allowed areas of the Bloch components for the trajectories introduced in Section \ref{sec:detell}.  While the deterministic ellipses pose outer boundaries for the Bloch components and retrodicted expectation values, they do not hold information on the actual distribution of trajectories realised over many experimental runs.

In Fig.\ \ref{fig:simulations} we show results of $10\,000$ Monte Carlo simulations of the SMEs \eqref{rhoeqn} and \eqref{EeqnNormalized} which allow sampling of the time dependent distribution of trajectories of the density and effect matrices of a monitored, decaying (pseudo)-spin as well as of the effective retrodicted Bloch vector components. As the $\rho$ and $E$ Bloch vector trajectories are confined to quite localized segments along the ellipses and they may be correlated with each other, the distribution of retrodicted Bloch vectors is restricted to more narrow regions than allowed by full deterministic ellipses. 

The dashed, black lines in Fig.\ \ref{fig:simulations} track the unconditional or ensemble averaged state, and it is seen that the deterministic ellipses of the conditional Bloch components of $\rho$ and $E$ does not include this state. This leads to a discrepancy between the most likely state represented by the bright, yellow areas in the color plots and the average state. This feature of the density matrix of a monitored quantum system is well-know, see e.g. \cite{webe14}. Similar results apply for the conditional trajectories of the effect matrix, and, as seen from Fig.~\ref{fig:simulations}b-c, when the qubit is post-selected in $\ket{\pm x}$, the most likely retrodicted set of expectation values differ from the unconditional retrodiction. 

\begin{figure}[]
 \begin{center}
\includegraphics[angle = 0, width = 0.48\textwidth]{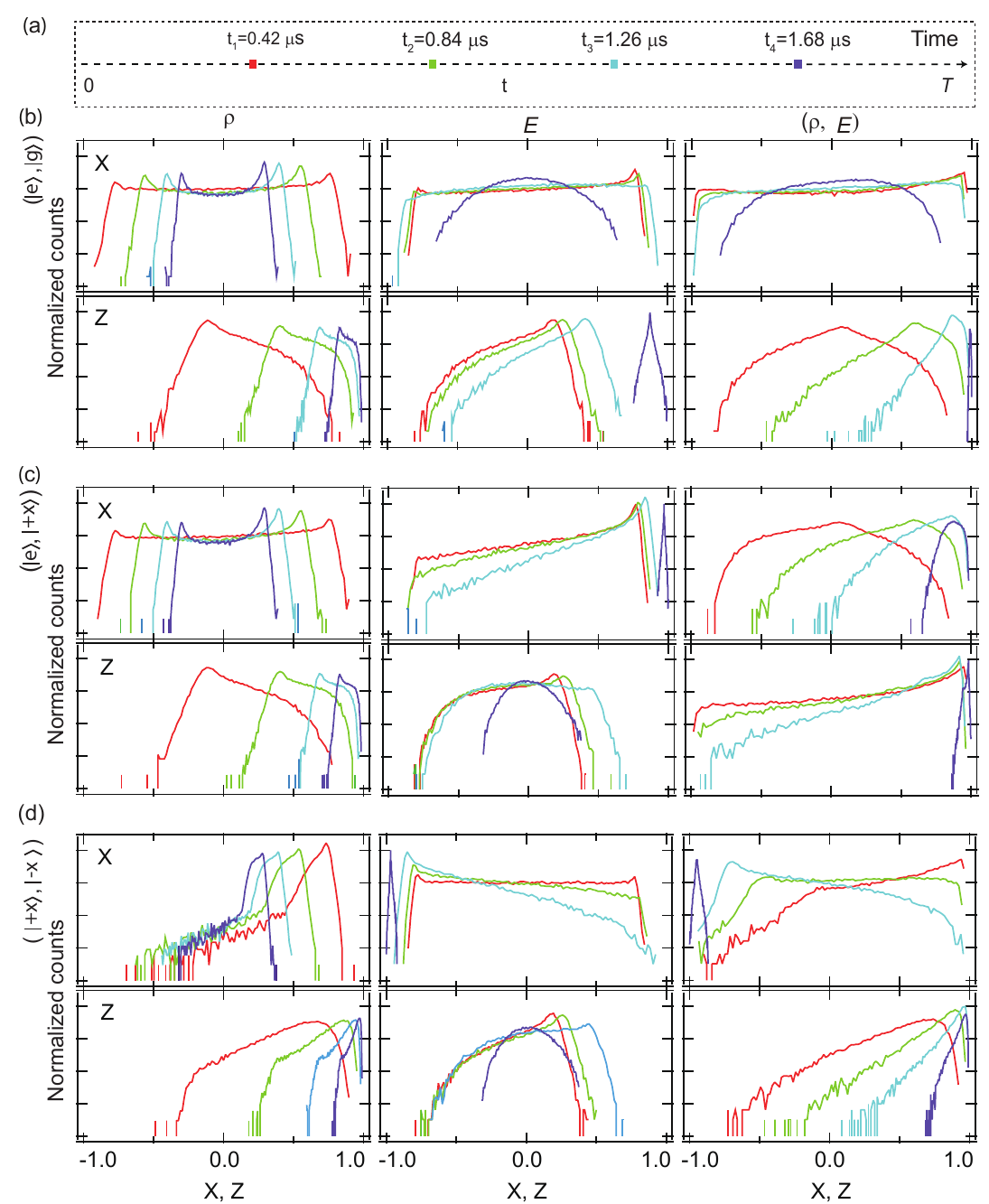}
\end{center}
\caption{Distribution of Bloch components at different times. (a), Time axis showing that at times $t=\{0.42,\ 0.84, \ 1.26, \ 1.68\}\ \mu$s, we calculate the distributions of $x$ and $z$ components. (b, c, d), Distribution of Bloch components $z$ and $z$ on a logarithmic scale based on $\rho$, $E$ and ($\rho$, $E$) respectively for pre- and post- selected states ($ |\psi_f\rangle$, $ |\psi_f\rangle$) at different times depicted by corresponding colors in Fig.\ \ref{fig_dis}a.  More than $3\times10^4$ successful pre- and post-selected runs out of $5\times10^6$ experimental repetitions were used to plot the distributions. } \label{fig_dis}
\end{figure}

The experiments similarly allow an analysis of the distribution of trajectories. In Figure~\ref{fig_dis}, we display separate histograms of the $x$ and $z$ Bloch components of $\rho$, $E$ and of the ($\rho$, $E$) retrodicted expectation values, corresponding to the different pre- and post-selections that were studied in Fig.~\ref{fig5}. These distributions agree with the theoretical simulations, and they confirm that the Bloch vector coordinates are restricted to finite intervals, and sometimes very well localized witin even tighter regions.

Both the simulations and the experiments provide distributions for the ($\rho$, $E$) trajectories that extend beyond the Bloch sphere and, e.g., approach the $x,z = \pm 1$ corner of the "Bloch square". We recall that the two coordinates of the retrodicted Bloch vector provide the probabilities of separate measurements of the $x$ and the $z$ pseudo spin components of the qubit. Close to $x,z = \pm 1$ we are thus able to make a confident, joint prediction for the outcome of a measurement of any of the two spin components. While this is normally forbidden by Heisenberg's uncertainty relation, we recall that we are not predicting the outcome of a future measurement, but rather retrodicting the outcome of a past one. If the state prior to such a past measurement is close to a $\sigma_z$ eigenstate (e.g., the ground state long time after preparation of the initial excited state), one could not have obtained the excited state in a $\sigma_z$ measurement. At the same time, if a subsequent final measurement yields $\sigma_x =1$, one could not possibly have measured  $\sigma_x=-1$ just prior to that. Hence, the majority  of Bloch components based on the ($\rho$, $E$) retrodiction may fall outside of the Bloch-sphere as seen in the third column of Fig.\ \ref{fig:simulations}b-c, and indicated by the curves in the third column of Fig.~\ref{fig_dis}c-d.


\begin{thebibliography}{29}
\expandafter\ifx\csname natexlab\endcsname\relax\def\natexlab#1{#1}\fi
\expandafter\ifx\csname bibnamefont\endcsname\relax
  \def\bibnamefont#1{#1}\fi
\expandafter\ifx\csname bibfnamefont\endcsname\relax
  \def\bibfnamefont#1{#1}\fi
\expandafter\ifx\csname citenamefont\endcsname\relax
  \def\citenamefont#1{#1}\fi
\expandafter\ifx\csname url\endcsname\relax
  \def\url#1{\texttt{#1}}\fi
\expandafter\ifx\csname urlprefix\endcsname\relax\def\urlprefix{URL }\fi
\providecommand{\bibinfo}[2]{#2}
\providecommand{\eprint}[2][]{\url{#2}}

\bibitem[{\citenamefont{Purcell}(1946)}]{purc46}
\bibinfo{author}{\bibfnamefont{E.~M.} \bibnamefont{Purcell}},
  \bibinfo{journal}{Phys. Rev.} \textbf{\bibinfo{volume}{69}},
  \bibinfo{pages}{681} (\bibinfo{year}{1946}).

\bibitem[{\citenamefont{Astafiev et~al.}(2010)\citenamefont{Astafiev, Zagoskin,
  Abdumalikov, Pashkin, Yamamoto, Inomata, Nakamura, and Tsai}}]{Asta10}
\bibinfo{author}{\bibfnamefont{O.}~\bibnamefont{Astafiev}},
  \bibinfo{author}{\bibfnamefont{A.~M.} \bibnamefont{Zagoskin}},
  \bibinfo{author}{\bibfnamefont{A.~A.} \bibnamefont{Abdumalikov}},
  \bibinfo{author}{\bibfnamefont{Y.~A.} \bibnamefont{Pashkin}},
  \bibinfo{author}{\bibfnamefont{T.}~\bibnamefont{Yamamoto}},
  \bibinfo{author}{\bibfnamefont{K.}~\bibnamefont{Inomata}},
  \bibinfo{author}{\bibfnamefont{Y.}~\bibnamefont{Nakamura}}, \bibnamefont{and}
  \bibinfo{author}{\bibfnamefont{J.~S.} \bibnamefont{Tsai}},
  \bibinfo{journal}{Science} \textbf{\bibinfo{volume}{327}},
  \bibinfo{pages}{840} (\bibinfo{year}{2010}), ISSN \bibinfo{issn}{0036-8075}.

\bibitem[{\citenamefont{Naghiloo
  et~al.}(2016{\natexlab{a}})\citenamefont{Naghiloo, Foroozani, Tan, Jadbabaie,
  and Murch}}]{nagh16}
\bibinfo{author}{\bibfnamefont{M.}~\bibnamefont{Naghiloo}},
  \bibinfo{author}{\bibfnamefont{N.}~\bibnamefont{Foroozani}},
  \bibinfo{author}{\bibfnamefont{D.}~\bibnamefont{Tan}},
  \bibinfo{author}{\bibfnamefont{A.}~\bibnamefont{Jadbabaie}},
  \bibnamefont{and} \bibinfo{author}{\bibfnamefont{K.~W.} \bibnamefont{Murch}},
  \bibinfo{journal}{Nature Communications} \textbf{\bibinfo{volume}{7}}
  (\bibinfo{year}{2016}{\natexlab{a}}).

\bibitem[{\citenamefont{Campagne-Ibarcq
  et~al.}(2014)\citenamefont{Campagne-Ibarcq, Bretheau, Flurin, Auff\`eves,
  Mallet, and Huard}}]{camp13}
\bibinfo{author}{\bibfnamefont{P.}~\bibnamefont{Campagne-Ibarcq}},
  \bibinfo{author}{\bibfnamefont{L.}~\bibnamefont{Bretheau}},
  \bibinfo{author}{\bibfnamefont{E.}~\bibnamefont{Flurin}},
  \bibinfo{author}{\bibfnamefont{A.}~\bibnamefont{Auff\`eves}},
  \bibinfo{author}{\bibfnamefont{F.}~\bibnamefont{Mallet}}, \bibnamefont{and}
  \bibinfo{author}{\bibfnamefont{B.}~\bibnamefont{Huard}},
  \bibinfo{journal}{Phys. Rev. Lett.} \textbf{\bibinfo{volume}{112}},
  \bibinfo{pages}{180402} (\bibinfo{year}{2014}).

\bibitem[{\citenamefont{Campagne-Ibarcq
  et~al.}(2016)\citenamefont{Campagne-Ibarcq, Six, Bretheau, Sarlette,
  Mirrahimi, Rouchon, and Huard}}]{camp16}
\bibinfo{author}{\bibfnamefont{P.}~\bibnamefont{Campagne-Ibarcq}},
  \bibinfo{author}{\bibfnamefont{P.}~\bibnamefont{Six}},
  \bibinfo{author}{\bibfnamefont{L.}~\bibnamefont{Bretheau}},
  \bibinfo{author}{\bibfnamefont{A.}~\bibnamefont{Sarlette}},
  \bibinfo{author}{\bibfnamefont{M.}~\bibnamefont{Mirrahimi}},
  \bibinfo{author}{\bibfnamefont{P.}~\bibnamefont{Rouchon}}, \bibnamefont{and}
  \bibinfo{author}{\bibfnamefont{B.}~\bibnamefont{Huard}},
  \bibinfo{journal}{Phys. Rev. X} \textbf{\bibinfo{volume}{6}},
  \bibinfo{pages}{011002} (\bibinfo{year}{2016}).

\bibitem[{\citenamefont{Steinberg}(1995)}]{Stein95}
\bibinfo{author}{\bibfnamefont{A.~M.} \bibnamefont{Steinberg}},
  \bibinfo{journal}{Phys. Rev. A} \textbf{\bibinfo{volume}{52}},
  \bibinfo{pages}{32} (\bibinfo{year}{1995}).

\bibitem[{\citenamefont{Aharonov et~al.}(2010)\citenamefont{Aharonov, Popescu,
  and Tollaksen}}]{ahar10}
\bibinfo{author}{\bibfnamefont{Y.}~\bibnamefont{Aharonov}},
  \bibinfo{author}{\bibfnamefont{S.}~\bibnamefont{Popescu}}, \bibnamefont{and}
  \bibinfo{author}{\bibfnamefont{J.}~\bibnamefont{Tollaksen}},
  \bibinfo{journal}{Physics Today} \textbf{\bibinfo{volume}{63}},
  \bibinfo{pages}{27} (\bibinfo{year}{2010}).

\bibitem[{\citenamefont{Aharonov et~al.}(2009)\citenamefont{Aharonov, Popescu,
  Tollaksen, and Vaidman}}]{ahar09}
\bibinfo{author}{\bibfnamefont{Y.}~\bibnamefont{Aharonov}},
  \bibinfo{author}{\bibfnamefont{S.}~\bibnamefont{Popescu}},
  \bibinfo{author}{\bibfnamefont{J.}~\bibnamefont{Tollaksen}},
  \bibnamefont{and} \bibinfo{author}{\bibfnamefont{L.}~\bibnamefont{Vaidman}},
  \bibinfo{journal}{Phys. Rev. A} \textbf{\bibinfo{volume}{79}},
  \bibinfo{pages}{052110} (\bibinfo{year}{2009}).

\bibitem[{\citenamefont{Aharonov et~al.}(1988)\citenamefont{Aharonov, Albert,
  and Vaidman}}]{ahar98}
\bibinfo{author}{\bibfnamefont{Y.}~\bibnamefont{Aharonov}},
  \bibinfo{author}{\bibfnamefont{D.~Z.} \bibnamefont{Albert}},
  \bibnamefont{and} \bibinfo{author}{\bibfnamefont{L.}~\bibnamefont{Vaidman}},
  \bibinfo{journal}{Phys. Rev. Lett.} \textbf{\bibinfo{volume}{60}},
  \bibinfo{pages}{1351} (\bibinfo{year}{1988}).

\bibitem[{\citenamefont{Aharonov and Vaidman}(1991)}]{ahar91}
\bibinfo{author}{\bibfnamefont{Y.}~\bibnamefont{Aharonov}} \bibnamefont{and}
  \bibinfo{author}{\bibfnamefont{L.}~\bibnamefont{Vaidman}},
  \bibinfo{journal}{Journal of Physics A: Mathematical and General}
  \textbf{\bibinfo{volume}{24}}, \bibinfo{pages}{2315} (\bibinfo{year}{1991}).

\bibitem[{\citenamefont{Tan et~al.}(2016)\citenamefont{Tan, Naghiloo,
  M\o{}lmer, and Murch}}]{tan16}
\bibinfo{author}{\bibfnamefont{D.}~\bibnamefont{Tan}},
  \bibinfo{author}{\bibfnamefont{M.}~\bibnamefont{Naghiloo}},
  \bibinfo{author}{\bibfnamefont{K.}~\bibnamefont{M\o{}lmer}},
  \bibnamefont{and} \bibinfo{author}{\bibfnamefont{K.~W.} \bibnamefont{Murch}},
  \bibinfo{journal}{Phys. Rev. A} \textbf{\bibinfo{volume}{94}},
  \bibinfo{pages}{050102} (\bibinfo{year}{2016}).

\bibitem[{\citenamefont{Tsang}(2009)}]{tsan09}
\bibinfo{author}{\bibfnamefont{M.}~\bibnamefont{Tsang}},
  \bibinfo{journal}{Phys. Rev. Lett.} \textbf{\bibinfo{volume}{102}},
  \bibinfo{pages}{250403} (\bibinfo{year}{2009}).

\bibitem[{\citenamefont{Guevara and Wiseman}(2015)}]{Guev15}
\bibinfo{author}{\bibfnamefont{I.}~\bibnamefont{Guevara}} \bibnamefont{and}
  \bibinfo{author}{\bibfnamefont{H.}~\bibnamefont{Wiseman}},
  \bibinfo{journal}{Phys. Rev. Lett.} \textbf{\bibinfo{volume}{115}},
  \bibinfo{pages}{180407} (\bibinfo{year}{2015}).

\bibitem[{\citenamefont{Rybarczyk et~al.}(2015)\citenamefont{Rybarczyk,
  Peaudecerf, Penasa, Gerlich, Julsgaard, M\o{}lmer, Gleyzes, Brune, Raimond,
  Haroche et~al.}}]{Ryba15}
\bibinfo{author}{\bibfnamefont{T.}~\bibnamefont{Rybarczyk}},
  \bibinfo{author}{\bibfnamefont{B.}~\bibnamefont{Peaudecerf}},
  \bibinfo{author}{\bibfnamefont{M.}~\bibnamefont{Penasa}},
  \bibinfo{author}{\bibfnamefont{S.}~\bibnamefont{Gerlich}},
  \bibinfo{author}{\bibfnamefont{B.}~\bibnamefont{Julsgaard}},
  \bibinfo{author}{\bibfnamefont{K.}~\bibnamefont{M\o{}lmer}},
  \bibinfo{author}{\bibfnamefont{S.}~\bibnamefont{Gleyzes}},
  \bibinfo{author}{\bibfnamefont{M.}~\bibnamefont{Brune}},
  \bibinfo{author}{\bibfnamefont{J.~M.} \bibnamefont{Raimond}},
  \bibinfo{author}{\bibfnamefont{S.}~\bibnamefont{Haroche}},
  \bibnamefont{et~al.}, \bibinfo{journal}{Phys. Rev. A}
  \textbf{\bibinfo{volume}{91}}, \bibinfo{pages}{062116}
  (\bibinfo{year}{2015}).

\bibitem[{\citenamefont{Gammelmark et~al.}(2013)\citenamefont{Gammelmark,
  Julsgaard, and M\o{}lmer}}]{gamm13}
\bibinfo{author}{\bibfnamefont{S.}~\bibnamefont{Gammelmark}},
  \bibinfo{author}{\bibfnamefont{B.}~\bibnamefont{Julsgaard}},
  \bibnamefont{and}
  \bibinfo{author}{\bibfnamefont{K.}~\bibnamefont{M\o{}lmer}},
  \bibinfo{journal}{Phys. Rev. Lett.} \textbf{\bibinfo{volume}{111}},
  \bibinfo{pages}{160401} (\bibinfo{year}{2013}).

\bibitem[{\citenamefont{Koch et~al.}(2007)\citenamefont{Koch, Yu, Gambetta,
  Houck, Schuster, Majer, Blais, Devoret, Girvin, and Schoelkopf}}]{koch07}
\bibinfo{author}{\bibfnamefont{J.}~\bibnamefont{Koch}},
  \bibinfo{author}{\bibfnamefont{T.~M.} \bibnamefont{Yu}},
  \bibinfo{author}{\bibfnamefont{J.}~\bibnamefont{Gambetta}},
  \bibinfo{author}{\bibfnamefont{A.~A.} \bibnamefont{Houck}},
  \bibinfo{author}{\bibfnamefont{D.~I.} \bibnamefont{Schuster}},
  \bibinfo{author}{\bibfnamefont{J.}~\bibnamefont{Majer}},
  \bibinfo{author}{\bibfnamefont{A.}~\bibnamefont{Blais}},
  \bibinfo{author}{\bibfnamefont{M.~H.} \bibnamefont{Devoret}},
  \bibinfo{author}{\bibfnamefont{S.~M.} \bibnamefont{Girvin}},
  \bibnamefont{and} \bibinfo{author}{\bibfnamefont{R.~J.}
  \bibnamefont{Schoelkopf}}, \bibinfo{journal}{Phys. Rev. A}
  \textbf{\bibinfo{volume}{76}}, \bibinfo{pages}{042319}
  (\bibinfo{year}{2007}).

\bibitem[{\citenamefont{Paik et~al.}(2011)\citenamefont{Paik, Schuster, Bishop,
  Kirchmair, Catelani, Sears, Johnson, Reagor, Frunzio, Glazman
  et~al.}}]{Paik11}
\bibinfo{author}{\bibfnamefont{H.}~\bibnamefont{Paik}},
  \bibinfo{author}{\bibfnamefont{D.~I.} \bibnamefont{Schuster}},
  \bibinfo{author}{\bibfnamefont{L.~S.} \bibnamefont{Bishop}},
  \bibinfo{author}{\bibfnamefont{G.}~\bibnamefont{Kirchmair}},
  \bibinfo{author}{\bibfnamefont{G.}~\bibnamefont{Catelani}},
  \bibinfo{author}{\bibfnamefont{A.~P.} \bibnamefont{Sears}},
  \bibinfo{author}{\bibfnamefont{B.~R.} \bibnamefont{Johnson}},
  \bibinfo{author}{\bibfnamefont{M.~J.} \bibnamefont{Reagor}},
  \bibinfo{author}{\bibfnamefont{L.}~\bibnamefont{Frunzio}},
  \bibinfo{author}{\bibfnamefont{L.~I.} \bibnamefont{Glazman}},
  \bibnamefont{et~al.}, \bibinfo{journal}{Phys. Rev. Lett.}
  \textbf{\bibinfo{volume}{107}}, \bibinfo{pages}{240501}
  (\bibinfo{year}{2011}).

\bibitem[{\citenamefont{Purcell et~al.}(1946)\citenamefont{Purcell, Torrey, and
  Pound}}]{purc46_2}
\bibinfo{author}{\bibfnamefont{E.~M.} \bibnamefont{Purcell}},
  \bibinfo{author}{\bibfnamefont{H.~C.} \bibnamefont{Torrey}},
  \bibnamefont{and} \bibinfo{author}{\bibfnamefont{R.~V.} \bibnamefont{Pound}},
  \bibinfo{journal}{Phys. Rev.} \textbf{\bibinfo{volume}{69}},
  \bibinfo{pages}{37} (\bibinfo{year}{1946}).

\bibitem[{\citenamefont{Wiseman and Milburn}(2010)}]{wisebook}
\bibinfo{author}{\bibfnamefont{H.}~\bibnamefont{Wiseman}} \bibnamefont{and}
  \bibinfo{author}{\bibfnamefont{G.}~\bibnamefont{Milburn}},
  \emph{\bibinfo{title}{Quantum Measurement and Control}}
  (\bibinfo{publisher}{Cambridge University Press}, \bibinfo{year}{2010}).

\bibitem[{\citenamefont{Jacobs and Steck}(2006)}]{jaco06}
\bibinfo{author}{\bibfnamefont{K.}~\bibnamefont{Jacobs}} \bibnamefont{and}
  \bibinfo{author}{\bibfnamefont{D.~A.} \bibnamefont{Steck}},
  \bibinfo{journal}{Contemp. Phys.} \textbf{\bibinfo{volume}{47}},
  \bibinfo{pages}{279} (\bibinfo{year}{2006}).

\bibitem[{\citenamefont{Jacobs}(2010)}]{Jacbook}
\bibinfo{author}{\bibfnamefont{K.}~\bibnamefont{Jacobs}},
  \emph{\bibinfo{title}{Stochastic Processes for Physicists: Understanding
  Noisy Systems}} (\bibinfo{publisher}{Cambridge University Press},
  \bibinfo{year}{2010}).

\bibitem[{\citenamefont{Murch et~al.}(2013)\citenamefont{Murch, Weber, Macklin,
  and Siddiqi}}]{murc13traj}
\bibinfo{author}{\bibfnamefont{K.~W.} \bibnamefont{Murch}},
  \bibinfo{author}{\bibfnamefont{S.~J.} \bibnamefont{Weber}},
  \bibinfo{author}{\bibfnamefont{C.}~\bibnamefont{Macklin}}, \bibnamefont{and}
  \bibinfo{author}{\bibfnamefont{I.}~\bibnamefont{Siddiqi}},
  \bibinfo{journal}{Nature} \textbf{\bibinfo{volume}{502}},
  \bibinfo{pages}{211} (\bibinfo{year}{2013}).

\bibitem[{\citenamefont{Weber et~al.}(2014)\citenamefont{Weber, Chantasri,
  Dressel, Jordan, Murch, and Siddiqi}}]{webe14}
\bibinfo{author}{\bibfnamefont{S.~J.} \bibnamefont{Weber}},
  \bibinfo{author}{\bibfnamefont{A.}~\bibnamefont{Chantasri}},
  \bibinfo{author}{\bibfnamefont{J.}~\bibnamefont{Dressel}},
  \bibinfo{author}{\bibfnamefont{A.~N.} \bibnamefont{Jordan}},
  \bibinfo{author}{\bibfnamefont{K.~W.} \bibnamefont{Murch}}, \bibnamefont{and}
  \bibinfo{author}{\bibfnamefont{I.}~\bibnamefont{Siddiqi}},
  \bibinfo{journal}{Nature} \textbf{\bibinfo{volume}{511}},
  \bibinfo{pages}{570} (\bibinfo{year}{2014}).

\bibitem[{\citenamefont{Tan et~al.}(2015)\citenamefont{Tan, Weber, Siddiqi,
  M\o{}lmer, and Murch}}]{Tan15}
\bibinfo{author}{\bibfnamefont{D.}~\bibnamefont{Tan}},
  \bibinfo{author}{\bibfnamefont{S.}~\bibnamefont{Weber}},
  \bibinfo{author}{\bibfnamefont{I.}~\bibnamefont{Siddiqi}},
  \bibinfo{author}{\bibfnamefont{K.}~\bibnamefont{M\o{}lmer}},
  \bibnamefont{and} \bibinfo{author}{\bibfnamefont{K.}~\bibnamefont{Murch}},
  \bibinfo{journal}{Phys. Rev. Lett.} \textbf{\bibinfo{volume}{114}},
  \bibinfo{pages}{090403} (\bibinfo{year}{2015}).

\bibitem[{\citenamefont{Bolund and M\o{}lmer}(2014)}]{Bolu14}
\bibinfo{author}{\bibfnamefont{A.}~\bibnamefont{Bolund}} \bibnamefont{and}
  \bibinfo{author}{\bibfnamefont{K.}~\bibnamefont{M\o{}lmer}},
  \bibinfo{journal}{Phys. Rev. A} \textbf{\bibinfo{volume}{89}},
  \bibinfo{pages}{023827} (\bibinfo{year}{2014}).

\bibitem[{\citenamefont{Dixon et~al.}(2009)\citenamefont{Dixon, Starling,
  Jordan, and Howell}}]{Dixon09}
\bibinfo{author}{\bibfnamefont{P.~B.} \bibnamefont{Dixon}},
  \bibinfo{author}{\bibfnamefont{D.~J.} \bibnamefont{Starling}},
  \bibinfo{author}{\bibfnamefont{A.~N.} \bibnamefont{Jordan}},
  \bibnamefont{and} \bibinfo{author}{\bibfnamefont{J.~C.}
  \bibnamefont{Howell}}, \bibinfo{journal}{Phys. Rev. Lett.}
  \textbf{\bibinfo{volume}{102}}, \bibinfo{pages}{173601}
  (\bibinfo{year}{2009}).

\bibitem[{\citenamefont{Pang et~al.}(2014)\citenamefont{Pang, Dressel, and
  Brun}}]{Pang14}
\bibinfo{author}{\bibfnamefont{S.}~\bibnamefont{Pang}},
  \bibinfo{author}{\bibfnamefont{J.}~\bibnamefont{Dressel}}, \bibnamefont{and}
  \bibinfo{author}{\bibfnamefont{T.~A.} \bibnamefont{Brun}},
  \bibinfo{journal}{Phys. Rev. Lett.} \textbf{\bibinfo{volume}{113}},
  \bibinfo{pages}{030401} (\bibinfo{year}{2014}).

\bibitem[{\citenamefont{Wiseman}(2002)}]{Wise02}
\bibinfo{author}{\bibfnamefont{H.~M.} \bibnamefont{Wiseman}},
  \bibinfo{journal}{Phys. Rev. A} \textbf{\bibinfo{volume}{65}},
  \bibinfo{pages}{032111} (\bibinfo{year}{2002}).

\bibitem[{\citenamefont{Naghiloo
  et~al.}(2016{\natexlab{b}})\citenamefont{Naghiloo, Tan, Harrington, K.,
  Lewalle, Jordan, and Murch}}]{caus16}
\bibinfo{author}{\bibfnamefont{M.}~\bibnamefont{Naghiloo}},
  \bibinfo{author}{\bibfnamefont{D.}~\bibnamefont{Tan}},
  \bibinfo{author}{\bibfnamefont{P.~M.} \bibnamefont{Harrington}},
  \bibinfo{author}{\bibnamefont{K.}},
  \bibinfo{author}{\bibfnamefont{P.}~\bibnamefont{Lewalle}},
  \bibinfo{author}{\bibfnamefont{A.~N.} \bibnamefont{Jordan}},
  \bibnamefont{and} \bibinfo{author}{\bibfnamefont{K.~W.} \bibnamefont{Murch}},
  \bibinfo{journal}{arXiv:1612.03189}  (\bibinfo{year}{2016}{\natexlab{b}}).

\end{thebibliography}
\end {document}